\documentclass[aps,twocolumn]{revtex4}
\usepackage{newlfont}
\usepackage{amssymb}
\usepackage{amsfonts}
\usepackage{amsmath}
\usepackage{graphicx}
\usepackage{bm}

\usepackage{graphicx}
\usepackage{epsfig}
\usepackage{newlfont}
\usepackage{amssymb}
\usepackage{amsfonts}
\usepackage{amsmath}
\usepackage{graphicx}
\usepackage{bm}

\begin{document}











\title{Cluster Entanglement Mean Field inspired approach to Criticality in Classical Systems}

\author{Aditi Sen(De) and Ujjwal Sen}

\affiliation{Harish-Chandra Research Institute, Chhatnag Road, Jhunsi, Allahabad 211 019, India}

\begin{abstract}

We propose an entanglement mean field theory inspired approach for dealing with interacting classical many-body systems. It involves a coarse-graining technique that 
terminates a step before the mean field theory: While mean field theory deals with only single-body physical parameters, the entanglement mean field 
theory deals with single- as well as two-body ones. We improve the theory to a cluster entanglement mean field, that deals 
with a fundamental unit of the lattice of the many-body system.
We apply these methods to interacting Ising spin systems in several lattice geometries and dimensions, and show that the predictions of the onset of 
criticality of these models are much better in the proposed theories as compared to the corresponding ones in mean field theories.

\end{abstract}

\maketitle

\section{Introduction and Main Results}

The search for bridges between many-body physics and quantum information has been very fruitful, and has led to many important discoveries and insights 
\cite{amaderAdP,Fazio-Osterloh-VedralRMP}. On the one hand, quantum information concepts have been used to  provide further tools to a 
many-body physicist, while on the other, realizable many-body physics systems ranging from quantum optics systems to ion traps are 
being tried as potential substrates for quantum information processing tasks.

It is rarely possible to treat an interacting many-body system exactly, and hence it is important to obtain approximate methods to 
deal with them. The mean field theory (MFT) \cite{MFT-book, Pathria, Mahan}, introduced by P. Weiss in 1907, 
is a very useful tool available to many-body physics to examine such systems, for 
both classical and quantum interacting many-body systems, with a low computational cost.

\textbf{Main thesis.} We propose that parallel to, but clearly different from, the MF class of theories \cite{MFT-book},
there exists an entanglement mean field class of theories to treat interacting classical many-body systems, 
that deals with one-body and two-body physical parameters in its 
self-consistency equations.

The mean field class of theories 
are 
an ultimate form of coarse-graining of the many-body system, in that it reduces the interacting 
many-body Hamiltonian to 
single body terms, and deals with single-body physical parameters in its self-consistency equstions \cite{MFT-book}. 
In contrast, the entanglement mean field class of theories proposes to stop a step before in the coarse-graining process, 
and reduces the parent Hamiltonian to a finite number of 
two-body terms, and deals with single- as well as two-body terms in its self-consistency equations.

The entanglement mean field class of theories provides us with a tool to go beyond the mean field class, and yet remain in the low-cost bracket. We believe that 
the formalism will be useful in 
investigation
and control of
many-body systems in several areas including condensed matter, ultra-cold gases, and quantum information.



An interesting 
improvement
of the mean field approach 
is the cluster mean field theory (CMFT) \cite{CMFT-review}, where one 
reduces the many-body system to a fundamental unit (``cluster'') of the many-body lattice, while still retaining footprints 
of the many-body parent as an undetermined parameter, although 
this undetermined parameter is 
still (like in MFT)
a one-site physical quantity (like, magnetization) of 
the many-body parent,
and it is determined by the self-consistency relation (CMF equation) equating the parameter to the same one-site quantity obtained from 
the cluster. We stress here that the self-consistency relations in \emph{both} MFT and CMFT are 
are based on
\emph{one}-site physical quantities, in particular, on magnetization.

To treat critical phenomena in interacting classical spin models inspired by entanglement mean field theory (EMFT), proposed 
for quantum systems in Ref. \cite{aaj-bristi-hoch-chhe},
one
reduces the many-body classical system to a two-body one 
while retaining imprints of the many-body 
parent as an undetermined parameter. In contrast to MFT, in EMFT, the undetermined parameter depends on a two-site physical 
quantity (like, two-point correlation) of the many-body parent. This parameter is then determined by the self-consistency 
relation (EMFT equation) equating, e.g. 
the 
two-point correlation of the many-body parent with that of the EMFT-reduced two-body system. 
Note that 
there is certainly no quantum entanglement \cite{HHHH-RMP} generated by applying the EMFT to classical spin systems.

We stress here that 
that the entanglement mean field theory is different from the cluster mean field approach. While the latter uses single-site 
physical parameters in its self-consistency equation, the former uses two-site ones. 
EMFT is also different from other useful techniques to deal with many-body systems, like the
renormalization group 
approaches \cite{egulo-renorm-boi}, with the latter using block decimation techniques on the whole lattice. These differences, both operational and result-wise, 
will be further underlined in Sec. \ref{ebar-baRi-jabo-sat-ta-chobbis-baje}.

In this paper, we also present a further improvement of
EMFT to a ``cluster EMFT'' (CEMFT) that reduces the many-body system to a fundamental unit of the 
many-body lattice, while retaining impressions of the  original many-body system as undetermined parameters. In contrast 
to CMFT, in CEMFT, the undetermined parameters depend \emph{both} on  one-site and two-site physical quantities (e.g., on magnetization and 
two-site correlation) of the many-body parent.  These parameters are then determined via \emph{coupled} self-consistency equations (CEMFT equations)
equating e.g. the magnetization of the original many-body system with that of the (CEMFT-reduced) cluster, and 
the 
two-point correlation of the many-body parent with that of the cluster.

Apart from CMFT, there are several other interesting generalizations of the mean field theory in the literature, including 
the Bethe-Peierls-Weiss approximation \cite{BPW}, 
the Onsager reaction field theory \cite{ORF}, 
the diagrammatic expansion method \cite{DEM}
the self-consistent correlated field theory \cite{SCCF}, 
the screened magnetic field theory \cite{SMF},
and
the correlated cluster mean field theory \cite{CCMFT}, to mention a few (see also \cite{MFT-review-kora-boi}). 
Improvements of the entanglement mean field theory in these directions are also possible, and 
will be pursued later.
Meanwhile, let us note here that all the above exciting 
examples, in the MF class of theories,
deal with single-body physical parameters 
in the respective self-consistency equations. 
In contrast, the EMF class of theories deal with single- as well as two-body physical parameters
in the EMFT class self-consistency equations.

\textbf{MFT vs. EMFT.}
Solving for 
magnetization and correlation functions 
from
the EMFT and CEMFT equations leads to the prediction of critical phenomena 
in the spin models. We apply the EMF and CEMF theories to the nearest-neighbor Ising model in one, two (hexagonal, square, and 
triangular), and three (cubic), dimensional lattices. The results are given in Table 1. In all the cases considered, in the different dimensions and 
geometries, EMFT gives better predictions over MFT, and CEMFT 
is better than 
CMFT. (Actually, EMFT is already better than CMFT 
in all the cases considered.) In the best case, EMFT is better than MFT by \(68\%\) 
and 
CEMFT is better than CMFT by 85\%, happening respectively for the hexagonal  and square lattice systems. In the worst case, 
EMFT is better than MFT by  42\%, and 
CEMFT is better than CMFT by 8\%, happening respectively for the triangular and cubic lattice systems.

\section{EMFT for Classical Models}
\label{ebar-baRi-jabo-sat-ta-chobbis-baje}


Before presenting the entanglement mean field theory inspired approach to classical spin models, let us briefly 
describe
the mean field theory
for such systems. 
Consider the nearest neighbor (classical) Ising model 
\begin{equation}
H= -J\sum_{\langle {\vec{i}}{\vec{j}} \rangle}\sigma_{\vec{i}} \sigma_{\vec{j}} 
\end{equation}
which represents a system of interacting (classical) spin-1/2 particles (Ising spins) on a \(d\)-dimensional lattice of an arbitrary fixed geometry.
The coupling strength \(J\) is positive, and \(\sigma_{\vec{i}} = \pm 1\) represents  the value of the Ising spin 
at the site \(\vec{i}\). 
\(\langle {\vec{i}} {\vec{j}}\rangle\) indicates that the corresponding sum runs over 
nearest neighbor lattice sites only.

The mean field theory consists in assuming that a particular spin, say at \(\vec{i_0}\), is special, and 
replacing all other spin operators by their mean values. 
Denoting the mean values of the spin operator \(\sigma_{\vec{i}}\) at the site \(\vec{i}\) by \(m\) (average magnetization), 
leads to an MFT Hamiltonian  \cite{MFT-book}, which we denote as 
\( H_{MFT} \).
One then solves the self-consistency equations (mean field equations)
\begin{equation}
m = \sum_{\mathcal{CF}(\mathcal{I})} \sigma \rho^\beta_{MFT}, 
\end{equation} 
for \(m\).
Here \(\rho^\beta_{MFT}\) is the mean field canonical equilibrium state 
\(\exp(-\beta H_{MFT})/Z_{MFT}\), \(Z_{MFT} = \sum_{\mathcal{CF}(\mathcal{I})}\exp(-\beta H_{MFT}))\) is the MF partition function,
\(\beta = \frac{1}{k_B T}\), with \(T\)
denoting temperature on the absolute scale, and \(k_B\)  the Boltzmann constant. 
Here, and in the rest of the paper, 
\(\mathcal{CF}(\mathcal{I})\) will denote all Ising configurations of all the spins involved in that particular case. 
In the MF equation as well as in the MF partition function, there is just a single spin left, and \(\mathcal{CF}(\mathcal{I})\) denotes the set of the two possibilities 
thereof. 
Substituting \(m\) in \(H_{MFT}\) and \(\rho^\beta_{MFT}\), one 
can
find the single-body physical properties of the system in the mean field limit.


\begin{figure}[h!]
\label{fig-chhobi-prothhom}
\begin{center}
\epsfig{figure=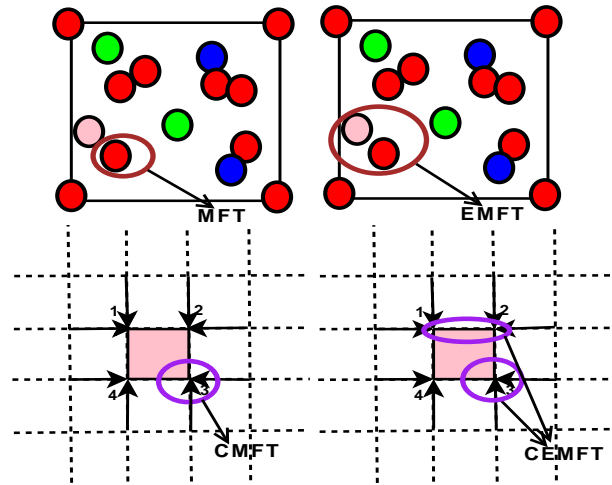, height=.27\textheight,width=0.45\textwidth}
\caption{(Color online) MF vs. EMF class of theories.
In MFT, a ``magnifying glass'' is put on a single particle of the many-body interacting system, and it leads to a self-consistency 
relation involving single-particle parameters. 
A different magnifying glass is employed in EMFT, which focusses attention on \emph{two} particles, and leads to a self-consistency relation 
involving \emph{two-particle physical parameters}.
Parallely, in CMFT, a fundamental unit (cluster) is chosen from the lattice which then is used to write self-consistency equations, again involving only 
single-site parameters. In CEMFT, the same cluster is used, but the self-consistency equations involve both one-site and two-site physical parameters.
}
\end{center}
\end{figure}

The entanglement mean field theory begins by noting that the square of an Ising spin random variable is unity. The two-body interaction Hamiltonian that we 
are dealing with, can be thought of an \(N\)-body interaction Hamiltonian (\(N\) being the total number of Ising spins in the system), in each term of which,
all but two random variables are constant (= unity). Let us call it a unit random variable. 
Since the square of any Ising random variable is unity, we can replace a unit random variable
on a site that is neighboring the nontrivial interacting spins of an interaction term,
by the square the Ising random variable at that site,
for all the  interaction 
terms in the Hamiltonian. 
Therefore, the term \(\sigma_{k-1,l}\sigma_{k,l}\) in a Hamiltonian on a two-dimensional square lattice can be replaced by 
\(\sigma_{k-1,l}\sigma_{k,l}\sigma_{k+1,l}\sigma_{k+1,l}\). The latter can be re-written as \(AB\), with
\(A=\sigma_{k,l}\sigma_{k+1,l}\), and \(B=\sigma_{k-1,l}\sigma_{k+1,l}\). Let us call this Hamiltonian as 
\(H_{EMFT}^{inter}\). For a given interaction term, let us call 
the site at which the replacement of the unit random variable by the square of the Ising random variable is done
as the dummy site for that interaction term. Given an interaction term, there can be several nearby sites that can act as the dummy site for that term.
So in  the case of the term \(\sigma_{k-1,l}\sigma_{k,l}\), \((k+1,l)\) is used as the dummy site.
We then assume 
%
%
%
that a certain \emph{pair} of two neighboring spins are ``special''. (See Fig. 1.)
Consider a \(d\)-dimensional lattice with coordination number \(\nu_{co}\).
There will be \(2(\nu_{co}-1)\) terms 
in the Hamiltonian \(H_{EMFT}^{inter}\), that will have the special pair, along with 
two more Ising spins in two  lattice sites (one of which, viz. the dummy site, is different from the special pair sites). 
In such a lattice, any one spin in the special pair is connected (via interactions in \(H\)) to \(\nu_{co} -1\) spins. Let this number 
(\(\nu_{co} -1\)) be denoted by \(\nu_E\), and be called the EMFT coordination number.  
We now
replace 
the non-special two-spin interactions (with nearby spins) in all the interaction terms in \(H_{EMFT}^{inter}\) by a 
constant multiple of their mean values \(C\). 
Physically, the mean value \(C\) represents the nearest neighbor correlation in the corresponding lattice. 
Since every interaction in \(H\) connecting to the special pair actually connects to \emph{one spin in the pair}, the non-special two-spin interactions in 
each term in \(H_{EMFT}^{inter}\) is 
replaced by \(\frac{1}{2}C\). 
The EMFT-reduced Hamiltonian, for the nearest neighbor Ising model on a \(d\)-dimensional lattice with EMFT coordination number \(\nu_E\),
will therefore be
\begin{equation}
\label{emft-asol}
{\cal H}_{EMFT} = -\frac{1}{2}J\nu_E C\sigma_{\vec{i}} \sigma_{\vec{j}}  
\end{equation}
where we have ignored the terms in the Hamiltonian which will not contribute to the EMFT equations below, and where 
we have assumed that the neighboring lattice sites \({\vec{i}}\) and \({\vec{j}}\) are special.
The self-consistency equation (EMFT equation) is
\begin{equation}
C = \sum_{\mathcal{CF}(\mathcal{I})}\sigma_{\vec{i}} \sigma_{\vec{j}} \varrho^\beta_{EMFT},
\end{equation}
for a system at temperature \(T\). 
Here \(\rho^\beta_{EMFT}\) is the entanglement mean field canonical equilibrium state 
\(\exp(-\beta H_{EMFT})/Z_{EMFT}\), \(Z_{EMFT} = \sum_{\mathcal{CF}(\mathcal{I})}\exp(-\beta H_{EMFT}))\) is the EMF partition function.
The EMFT equation  is to be solved 
for \(C\)
for obtaining the two-particle physical properties of 
\(H\) in the EMFT limit.
In a typical situation, there is a finite temperature \(T_c\), the critical temperature, 
that depends on the lattice geometry and dimension, above which 
the EMFT equation provides a nontrivial (i.e. nonzero) solution.

The values of the critical temperatures have been obtained for the interacting Ising systems in different dimensions and geometries, 
and are given in Table 1. The predictions of the EMF theory are always better than the corresponding ones from MFT in the cases considered, 
and almost always more than 50 percent better,
and for the hexagonal (honeycomb) lattice in two dimesnions, EMFT is 68\% better than MFT. See Table 1 for 
further details. 


In the entanglement mean field theory, the nature of the interactions propagating in the lattice geometry, enters the predictions through the 
the existence of an interaction term (\(\sigma_{\vec{i}} \sigma_{\vec{j}}\)) and 
a two-site physical parameter (the mean correlation \(C\)) in the EMFT Hamiltonian \(H_{EMFT}\), and their interplay in the self-consistency equation (EMFT equation). 
These features are absent in MFT, where the MFT Hamiltonian contains a single Ising random variable and a single-site physical parameter.
Moreover, an EMFT coordination number enters the stage in EMFT, while it is the coordination number in MFT. These differences 
lead to the better memory of the EMFT of its many-body parent, and a consequent better performance of the EMFT over MFT.

Note that the EMFT approach is different from the cluster MFT \cite{CMFT-review} as there the fundamental unit changes 
with the lattice geometry and dimension, while here
we always work with two special spins regardless of the lattice geometry. 
More importantly, the cluster MFT still (i.e., as in MFT) uses single-site properties to 
construct the self-consistency equations, while in EMFT, we use two-site properties to do the same. 
Additionally, both MFT and CMFT leads to effective Hamiltonians consisting of either a single random varible or a sum over single random variables, while
in EMFT, we deal with effective Hamiltonians that retain interaction terms involving two random variables. 
These are some of the operational differences. Result-wise,
a glance at Table 1 reveals that the predictions, for the critical 
temperatures of the different models, of EMFT and CMFT, are very different. Indeed, in all the cases considered, EMFT performs better than CMFT, with the 
prediction in the case of the three-dimensional cubic lattice being 49\% better.  
It is possible to obtain a generalization of the EMFT 
approach, towards a ``cluster EMFT'', for a better consideration of the lattice geometry, and we do so in the succeeding section.

The EMF approach is also different from the other techniques to handle many-body models. In particular, it is unlike the
renormalization group approach \cite{egulo-renorm-boi}, where block decimation techniques are used on the whole lattice, 
and free energy of the decimated lattice is equated to that of the original. Result-wise, application of the renormalization group to, e.g. the nearest 
neighbor Ising model on the square lattice predicts a critical temperature (in units of \(k_B T/J\)) at \(2.55\) \cite{Pathria}. 
The EMFT prediction is \(3\). The CEMFT prediction (considered below) is \(2.08\), with the exact value being \(2.27\) \cite{exact-values} 

A similar formalism as above works (for both EMFT and cluster EMFT),  with  suitable modifications, for quantum spins, higher discrete spins, continuous 
spins, more complex lattices and 
interactions, etc.
Also, both the mean field theory as well as the EMFT has been described for the ferromagnetic cases. The antiferromagnetic 
case requires some modifications in the mean field theory, and correspondingly some changes in the EMFT (and CEMFT). These will not be 
discussed in this paper.


\section{Cluster EMFT for Classical Models}
\label{teRe-khide-peye-gyachhe}

As has been noted before, the interactions of the many-body parent propagating in the lattice, 
are taken care of  in the entanglement mean field theory, by the EMFT coordination number, and the interplay of the mean correlation \(C\) and 
the interaction term in \(H_{EMFT}\) in the self-consistency equation. We have seen that this gives a better 
consideration to the interactions in the parent Hamiltonian than that in MFT. 
Towards  improving our approximations, we now include the lattice structure along with interactions between spin variables. 
%
We call it the cluster 
entanglement mean field theory, and is described as follows. 

For definiteness, consider a two-dimensional square lattice. See Fig. 1. A cluster in this case is a fundamental square of four spins. 
Let us focus on a particular cluster of four spins. 
These four spins interact among themselves
by four interaction terms in \(H\). They are the intra-cluster interactions. 
This basic unit, consisting of  four spins, also interact with other spins in neighboring 
clusters via inter-cluster interaction terms in \(H\). We first consider the intra-cluster terms. 
Just like in the case of the entanglement mean field theory, in every intra-cluster interaction term, 
we replace a unit random variable at a nearby dummy site by the square of an Ising random variable at that site. The difference is that 
the dummy site is now always chosen from the among the sites in the chosen cluster. This is just like in CMFT, where
only the intra-cluster spins are involved in producing the terms of the form \(m\sigma_{\vec{i}}\). 
So for the closen square cluster consisting of the sites 
formed by rows \(k, k-1\) and columns \(l,l-1\), for the intra-cluster term
\(\sigma_{k-1,l}\sigma_{k,l}\) in the Hamiltonian \(H\), 
the site \((k,l-1)\) can act as a dummy site, whereby we obtain the term
\(\sigma_{k-1,l}\sigma_{k,l}\sigma_{k,l-1}\sigma_{k,l-1}\). Similarly as in EMFT, the so-obtained term 
can be re-written as \(A_\chi B_\chi\), with
\(A_\chi=\sigma_{k,l}\sigma_{k,l-1}\), and \(B_\chi=\sigma_{k-1,l}\sigma_{k,l-1}\).
We now replace \(B_\chi\) by the unit multiple of the mean value \(C\), and consequently, the contribution of this intra-cluster term to the cluster EMFT Hamiltonian
\(\mathcal{H}_{CEMFT}\) is \(-JC\sigma_{k,l}\sigma_{k,l-1}\). 
The entanglement coordination  number of EMFT is absent in CEMFT, as the latter itself depends on the lattice geometry, and hence strenthens the 
approximation.
The inter-cluster terms are taken care of by replacing them with effective fields at the corresponding spins of the chosen cluster, and 
these terms  are exactly the
same as in CMFT. The terms in \(H\) that are neither intra- not inter-cluster, do not appear 
in the considerations below, and are therefore ignored. Therefore, for a two-dimensional square lattice, with the chosen cluster, the cluster EMFT Hamiltonian is 
\begin{equation}
 \mathcal{H}_{CEMFT}^{sq} = -JC \sum_{\langle {\vec{i}}{\vec{j}} \rangle_\chi} \sigma_{\vec{i}} \sigma_{\vec{j}} 
                                -2Jm \sum_{\vec{i}_\chi} \sigma_{\vec{i}}
\end{equation}
where the first sum 
runs over nearest neighbor sites of the chosen cluster, and the second runs over sites of the same. The factor 2 in the second term comes from 
the fact that  we are considering a 
square lattice, so that every spin in the chosen cluster is connected (via an interaction term in \(H\)) to \emph{two} spins in the neighboring clusters.
Here, \(C\) denotes the nearest neighbor correlation of the lattice under consideration, and \(m\) the corresponding magnetization. 
One may similarly find the cluster EMFT Hamiltonian \(H_{CEMFT}\) for other models.

At this point, both \(C\) and \(m\) are undetermined. They are to be solved from the self-consistency relations (CEMFT equations)
equating the correlation \emph{and} the magnetization of the chosen cluster with the corresponding ones of the whole lattice: 
\begin{eqnarray}
C=  \sum_{\mathcal{CF}(\mathcal{I})}\sigma_{\vec{i}} \sigma_{\vec{j}} \varrho^\beta_{CEMFT}, \nonumber \\
m= \sum_{\mathcal{CF}(\mathcal{I})}\sigma_{\vec{k}}  \varrho^\beta_{CEMFT},
\end{eqnarray}
where \(\vec{i}\) and \(\vec{j}\) are any two nearest neighbor sites, and \(\vec{k}\) a particular site, in the chosen cluster.
Here \(\rho^\beta_{CEMFT}\) is the cluster entanglement mean field canonical equilibrium state 
\(\exp(-\beta H_{CEMFT})/Z_{CEMFT}\), \(Z_{CEMFT} = \sum_{\mathcal{CF}(\mathcal{I})}\exp(-\beta H_{CEMFT}))\) is the CEMF partition function.
The CEMFT equations form a set of coupled self-consistency relations for \(C\) and \(m\), and their nontrivial solution set exists only after a certain temperature, 
which is the critical temperature obtained from the cluster entanglement mean field theory.

The table below gives the predictions for the critical temperatures for the nearest neighbor Ising model in different
lattice geometries and dimensions.  In this paper, we have obtained the predictions from entanglement mean field theory and 
cluster entanglement mean field theory. The predictions from mean field theory can be obtained, e.g. from Refs. \cite{MFT-book}. 
The predictions from cluster mean field theory are obtained in Refs. \cite{CMFT-review} and references therein. 
The exact and series results are obtained in Refs. \cite{exact-values} and references therein.
\begin{widetext}
\begin{equation}
\begin{array}{|c||c|c|c||c|c|c||c|}
\hline
\mbox{Lattice} & \mbox{MFT} & \mbox{EMFT} & \mbox{Improvement (\%)} & \mbox{CMFT} & \mbox{CEMFT} & \mbox{Improvement (\%)} & \mbox{Exact/Series} \\
\hline
\mbox{Linear} & 2 & 1 & 50 & 1.28 & 1.05 & 17.97 & 0 \\
\mbox{Hexagonal} & 3 & 2 & 67.57 & 2.335 & 1.21  & 63.53 & 1.52\\
\mbox{Square} & 4 & 3 & 57.8 & 3.5 & 2.08 & 84.55 & 2.27\\
\mbox{Triangular} & 6 & 5 & 42.37 & 5.64 & 3.99 & 82.5 & 3.64\\
\mbox{Cubic} & 6 & 5 & 66.67 & 5.49 & 3.61 & 8.16 & 4.51\\
\hline
\end{array}
\nonumber
\end{equation}
\begin{center}
Table 1: 
A comparison of the critical temperatures obtained for the nearest neighbor Ising model in different lattices and geometries. 
Except for those in the two columns that mention the improvements, the numbers in the other columns are in units of 
\(k_B T/J\). There are two columns with the heading ``Improvement'', of which the left one shows the improvement in the 
EMFT prediction over that from MFT, while the right one shows that for CEMFT over CMFT. 
\end{center}
\end{widetext}

\section{Conclusion}

We have proposed an entanglement mean field theory for dealing with classical interacting many-body models. Distinct from the mean field approach to 
interacting systems, the entanglement mean field one reduces the many-body parent Hamiltonian into a two-body one involving undetermined mean values of 
two-site physical parameters of the many-body parent. These undetermined parameters are determined via self-consistency equations between mean values of 
the two-body physical quantity of the 
reduced Hamiltonian and the many-body parent. We then generalize the concept to a cluster entanglement mean field theory where we work with a 
fundamental unit of the lattice. The self-consistency relations in this case are a set of coupled equations of single-site and two-site 
physical quantities. Solving these self-consistency equations lead to the predictions of critical temperatures of the models considered, which we then
compare with the previous results. In all the cases considered, in the different geometries and dimensions, the predictions of the entanglement mean
field theory are better than mean field theory (68\% at most, and  42\% at least), 
and the same of the cluster entanglement mean field theory are better than cluster mean field theory (85\% at most, and 8\% at least).

\end{document}